\documentclass{nsart_eng}
\usepackage{cite}
\usepackage[dvips]{graphicx}
\usepackage{amsmath,amssymb}

\year{2017} \volume{8} \nomer{1} \firstpage{1}

\let\le\leqslant

\let\ge\geqslant

\begin{document}

\setcounter{page}{1}
\newcommand{\re}[1]{(\ref{#1})}
\newcommand{\lab}[1]{\label{#1}}
\newcommand{\ci}[1]{\cite{#1}}
\renewcommand{\baselinestretch}{1.25}
\newcommand{\bfr}{\begin{flushright}}
\newcommand{\bfl}{\begin{flushleft}}
\newcommand{\efl}{\end{flushleft}}
\newcommand{\efr}{\end{flushright}}
\newcommand{\bc}{\begin{center}}
\newcommand{\ec}{\end{center}}
\newcommand{\be}{\begin{equation}}
\newcommand{\ee}{\end{equation}}
\newcommand{\bea}{\begin{eqnarray}}
\newcommand{\eea}{\end{eqnarray}}
\newcommand{\ba}{\begin{array}}
\newcommand{\ea}{\end{array}}
\newcommand{\edc}{\end{document}}
\newcommand{\ul}{\underline}
\newcommand{\ri}{\rightarrow\infty}
\newcommand{\li}{\leftarrow\infty}
\newcommand{\ra}{\rightarrow}
\newcommand{\la}{\leftarrow}
\newcommand{\ds}{\displaystyle}
\newcommand{\dsf}{\displaystyle\frac}
\newcommand{\dt}{\Delta{t}}
\newcommand{\il}{\int\limits}
\newcommand{\pal}{\partial}
\newcommand{\cal}{\mathcal}
\newcommand{\bone}{{\bf 1}}
\newcommand{\gComment}[1]{}
\renewcommand{\gComment}[1]{\textcolor{red}{Gerardo: #1}}


\title[{\it Normal state pair nematicity and hidden magnetic order...}]
{Normal state pair nematicity and hidden magnetic order and metal-insulator (fermion-boson)- crossover 
origin of pseudogap phase of cuprates II}

\author[{\it B.~Abdullaev, D.\,B.~Abdullaev, C.\,-H.~Park, M.\,M. Musakhanov}]
{B.~Abdullaev$^1$, D.\,B.~Abdullaev$^1$, C.\,-H.~Park$^2$, M.\,M. Musakhanov$^3$}

\address{ 
$^1$Institute of Applied Physics, National University of Uzbekistan,
Tashkent 100174, Uzbekistan\\
$^2$Research Center for Dielectric and Advanced Matter Physics,
Department of Physics, Pusan National University, 30
Jangjeon-dong, Geumjeong-gu, Busan 609-735, Korea\\
$^3$National University of Uzbekistan,
Tashkent 100174, Uzbekistan}

\email{bakhodir.abdullaeff@yandex.ru, cpark@pusan.ac.kr, yousufmm@list.ru}












\begin{abstract}
In the present paper II, we will gain an understanding of the nematicity, 
insulating ground state (IGS), nematicity to stripe phase transition, Fermi pockets evolution, and
resistivity temperature upturn, as to be metal - insulator (fermion-boson)- crossover (MIC) 
phenomena for the pseudogap (PG) region of cuprates. While in the paper I~\ci{Abdullaev0}, we 
obtained an understanding 
of the observed heat conductivity downturn, anomalous Lorentz ratio, insulator resistivity 
boundary, nonlinear entropy as manifestations of the same MIC. The recently observed nematicity and hidden 
magnetic order are related to the PG pair intra charge and spin fluctuations. We will try to 
obtain an answer on the question: why ground state of YBCO is Fermi liquid oscillating and of 
Bi-2212 is insulating? We will also clarify the physics of the recently observed MIC results 
of Laliberté et al.~\ci{Laliberté} and explain the long-discussed
transition of the electric charge density from doping to doping+1 dependence at the critical 
doping. We predict that at the upturns this density should have the temperature dependence
$n\sim T^3n_2$ for $T\rightarrow 0$, where $n_2$ is density for dopings close to the critical value. 
We understood that the upturns before and after the first critical doping have the same nature. 
We will find understanding of all above mentioned phenomena within the PG pair physics. 
\end{abstract}


\maketitle




\section{Introduction}
\label{sec1}
Information about the normal state raduis of localization of each electron or hole, or its wave 
function, may play an important role for high temperature $T_c$ cuprate superconductors.
The size of this raduis may lead to an understanding of the physics of, for example, nematicity 
and stripe phases. The minimal size for square form nanoregions (NRs) (or of pseudogap 
(PG) and high $T_c$ superconductivity (HTS) pairs), measured recently in the scanning 
tunneling microscopy (STM) experiments of Gomes et al.~\ci{Gomes} and Pan et al.~\ci{Pan} 
for visualization of the energy gap formation in the cuprate $Bi_2Sr_2CaCu_2O_{8+\delta}$, 
has varied between nearly 4.5 $a$ and 2.6 $a$ of the $a-b$ plane crystal spacing constant $a$, 
when doping varies between two critical values in a doping-temperature phase diagram. Electric
charge analysis of these NRs, made by Abdullaev et al.~\ci{Abdullaev1}, reveals that at the first 
critical doping we have one boson inside of each NR, while at the second critical doping, 
one boson and one fermion. A wide spatial spreading of particle-charge in the underdoped 
region indicates that the role of the dielectric parent compound of cuprates is unimportant, which
leads to universality of the properties of all underdoped copper-oxides  with the same $a$.

The concept of Landau Fermi-liquid quasi-particles (QPs) plays a central role in the 
understanding of the normal state physics for conventional electric conducting 
materials~\ci{ashcroft,abrikosov}. Typically, the energies of these QPs 
depend on momentum and they appear close to the Fermi surface. 

For the HTS cuprates (copper-oxides), the normal PG state exhibits highly non-Fermi 
liquid properties (see, for example, our papers: I~\ci{Abdullaev0} and~\ci{Abdullaev01}, 
for a comprehensive list of references) and it is unclear whether QPs defined 
in the momentum space are essential for the PG physics or not. However, as it will be shown 
in this paper, the real space PG pairs mentioned above and observed in the STM experiment are
essential for an understanding of this physics.

In our first paper I~\ci{Abdullaev0} we gained an understanding of the following observed MIC 
phenomena: heat conductivity downturn, anomalous Lorentz ratio, insulator resistivity boundary, 
and nonlinear entropy. In the present paper (II), we will try to understand of other observed MIC phenomena: 
IGS, nematicity- and stripe-phases, Fermi pockets evolution, and resistivity temperature upturn.
A question can appear about the relation of Fermi pockets evolution to the MIC phenomenon.
This question is experimentally studied in detail in the Ref.~\ci{Collignon} for 
$La_{1.6\, -\, x} Nd_{0.4}Sr_xCuO_4$ copper-oxide. However, the origin of Fermi pockets
and what is origin of their evolution with doping has still not been adequately explained in the literature.
We will see below that the disconnected Fermi pockets may just be indication of a (fermion) 
IGS and this state's evolution with doping coincides with these pockets evolution, 
while the real origin of the MIC may be bosonic.

The seminal STM experiments of Gomes et al.~\ci{Gomes} and Pan et al.~\ci{Pan} on the visualization 
of the real space PG and HTS NRs, which exhibited an energy gap, together with their electric 
charge and percolation analysis, made by Abdullaev et al.~\ci{Abdullaev1}, have deeply extended our
understanding of the HTS cuprate physics. The theoretical calculation, shown in the Ref.~\ci{Abdullaev1}, 
has further justified the possible existence of single bosons in two-dimensional systems as $a-b$ 
planes of cuprates. The QPs, PG and HTS pairs-single bosons, have become the fundamental particles
in our semi-phenomenological Coulomb single boson and single fermion two liquid model, which positions
are formulated in the following list~\ci{Abdullaev01}.

1. The doping charges, in the form of individual 
NRs, are embedded in the insulating parent compound of HTS copper oxides.
2. Before the first critical doping $x_{c1}$ with NR size $\xi_{coh}=17 \AA$, 
they are not percolated single bosons. 3. The origin of single bosons is in 
the anyon bosonization of 2D fermions. 4. At the first critical doping level $x_{c1}$,
the percolation of single boson NRs and thus HTS appears; there appear also from
$x_{c1}$ doping single fermions, but up to second critical doping level, $x_{c2}$, their
NRs do not percolate, thus single fermions between $x_{c1}$ and $x_{c2}$ are 
insulating. 5. The value $x_{c1}=0.05$ is universal for all copper oxide
HTSs, since percolating single boson NRs cover $50\%$ of the 2D sample area (like 
connecting squares in a chessboard); the same situation takes place with NRs for fermions 
at $x_{c2}$. 6. The normal phase charge conductivity appears from $x_{c2}$ at 
$T=0$ or above PG temperature boundary $T=T^*$, where the percolation of 
single fermions appears, while for temperatures between $T_c$ and $T^*$, there 
exists (fermion-boson) MIC. 7. The spatially rare charge density object, single 
boson, with NR size between $\xi_{coh}=17 \AA$ and $\xi_{coh}=10 \AA$, which 
correspond to $x_{c1}$ and $x_{c2}$ dopings, has zero total but fluctuating 
inside of NR spin (this rareness leads also to fluctuating charge inside of NR). 
8. The increase of boson spin fluctuations with doping or 
temperature results in a transition of bosons into fermions, which occurs at
PG $T^*$ or at $x_{c2}$. 9. At zero external magnetic field, the HTS is a result 
of the Bose-Einstein condensate of single bosons. 10. At high external magnetic 
field strengths, the PG insulating ground state is the result of a plasmon gas consisting of these bosons.
We note here that $a\approx 3.8 \AA$.

The majority of these positions are the result of experimental data analysis. Below, when we
will discuss positions from this list, we refer to experimental papers, which confirm
the validity of these positions, or use the physical arguments for their support. The background 
of our model, existence of single bosons, has been derived in our microscopical anyon 
calculation and found its confirmation in the Gomes et al. and Pan et al.~\ci{Gomes,Pan} STM experiments. 
Results of this analysis together with microscopical single boson picture 
have been combined in the Coulomb two liquid model. 

The first two points of these positions have been supported by the experimental 
paper~\ci{Mclaughlin}, authors of which reported on the emergent transition 
for superconducting fluctuations in the deep antiferromagnetic phase at a 
remarkably low critical doping level, $x_c= 0.0084$, for ruthenocuprates, 
$RuSr_2(R,Ce)_2Cu_2O_{10-\delta}$ with $R = Gd,Sm,Nd$. In this paper, it was claimed 
that those fluctuations have an intrinsic electronically-inhomogenous nature and 
provide new support for bosonic models of the superconducting mechanism.

In the introduction section of Ref.~\ci{Abdullaev01}, we have described a qualitative
understanding of the cuprate physics within our Coulomb single boson and single fermion two 
liquid model. The present paper will be devoted to application of PG pairs for understanding 
of MIC phenomena listed above. In the Sec.~\ref{sec2} we will give an understanding
of the nematicity, IGS, nematicity-stripe phases transition, and Fermi pockets evolution 
with doping. We explain in detail the physical origin of each phenomenon using the
NRs concept. To understand of the resistivity temperature upturn effect, we will 
use the Sec.~\ref{sec3}. In the Sec.~\ref{sec4} we will discuss about the nematicity and
hidden magnetic order as manifestations of the intra PG pair charge and spin fluctuations.
We argue in this section why these fluctuations may be an indication of the PG pairs.
In the Sec.~\ref{sec5}, we formulate an interesting question: "Why is the ground state of 
$YBCO$ an oscillating Fermi liquid and for $Bi-2212$ is insulating?" This is taken from the experimental
analysis and we attempt to answer to it. For interpretation of the recently published experimental 
MIC results, from the point of view of our Coulomb two liquid model, we will devote the 
Sec.~\ref{sec6}. And we conclude our paper with the Sec.~\ref{sec7}.

\section{Nematicity, IGS, Nematicity-stripe phases transition, and Fermi pockets evolution}
\label{sec2}  
The definition of the intra-unit-cell nematicity order in the copper-oxides is, probably,  
given in the Ref.~\ci{Lawler} and it means the breaking of rotational $C_4$ symmetry by the 
electronic structure within $CuO_2$ unit cell, while the translational symmetry over all cells 
is conserved. The elemental structure for cuprate's unit cell has square form with $C_4$ 
rotational symmetry. Authors of Ref.~\ci{Lawler} have detected in the STM experiment the 
prominent electronic structure (charge distribution) along of some sides of this cell, in 
which the two closest oxygen atoms $O_2$ to the $Cu$ atom played an important role, for the 
underdoped $Bi_2Sr_2CaCu_2O_{8+\delta}$.

However, for underdoped cuprate the area of NR is largest. Thus, a single boson unit 
charge is spreaded over the whole NR, which covers several unit cells, giving a very
low regional charge density of the NR. The connection of $Cu$ and $O$ atoms in the unit 
cell of parent compound is electrically polarized. It redistributes and fixes a charge 
inside of the NR. As a result, a pinning of the NR occurs. The pattern of NRs with random 
pinning of their charges in real space yields the pattern for the electronic nematicity.
On the other hand, the soft pinning of NR charges does not destroy the system's translational
symmetry.
 
The pinning of single boson charges in a strong magnetic field (we remember that this
magnetic field destroys the HTS and experiment detects IGS (see our paper I)) by the cuprate 
parent compound is similar to the Wigner crystal of 2D electrons, pinning in a disorder potential and 
in the same magnetic field. However, this Wigner crystal, under the same external potentials,
results in the IGS of electrons (see Refs.~\ci{Drichko,Shayegan}). Therefore, the plasmon gas of 
pinned 2D charged single bosons has the IGS. Thus, if the pinned charged bosons are insulating,
then the plasmons of this gas is also insulating. The last leads to the IGS of bosonic
insulator or to IGS of cuprates (see Refs.~\ci{Ando,Fournier,Ono,Ando1,Boebinger} and 
point 10. in the above list of our model positions). This can be detected by measuring of 
this plasmon gas frequency using method described in~\ci{Drichko,Shayegan}. Then, the bosonic 
insulator - bosonic metal transition can experimentally be observed at a point for this frequency 
when it goes to zero. Physically, this transition corresponds to the changing of the charged boson
gas elementary excitation spectrum from a plasmon part into a free particle one~\ci{Abdullaev0}.

Authors of Ref.~\ci{Caprara} have displayed in their Fig. 8 the schematic evolution of the 
nematic and stripe charge order phases with doping. They showed that a nematic 
charge order, with violation of the system's rotational symmetry and small spatial size of this order, 
dominates at initial dopings of the HTS dome phase diagram. While the stripe phase, with violation of 
the system's translational symmetry, appears for dopings close to the optimal (critical) level.

The fact that the system's translational symmetry is violated means that a factor, which violates 
this symmetry, has the same spatial size as $CuO_2$ unit cell. With an increase of doping in the
interval between two critical levels and close to the optimal one, the spatial sizes of NRs 
decrease and single bosons transform into single fermions~\ci{Abdullaev1}. Together with 
insulating single fermions, appearing at the first critical doping (see point 4. of the 
model positions), the concentration of fermions becomes dominant in the mixture of
bosons and fermions. Therefore, for violation of the system's translational symmetry, and 
thus for a stripe phase, are responsible the NRs occupied by charged single fermions (the 
real space size of the fermion NR is close to $a$). 

The schematic evolution of the nematic and stripe charge order phases with doping resembles 
the bosonic insulator - bosonic metal transition crossover. The temperature-doping boundary
of both these evolutions is desribed by formula (see Eq. 13 of Ref.~\ci{Abdullaev0}):
\be T_{bIM} \approx 0.368 
[x(1-x/x_c)]^{2/3} Ry \, , \lab{wf7a} \ee where $x_c$ and $Ry$ are the critical doping and
Rydberg atomic energy unit, respectively. 

Now, it is worthwhile to discuss about Fermi pockets evolution with doping as an 
additional signature of the MIC. Fermi pockets and stripe phases belong to a very controversial subject 
of PG cuprate physics. There is a widely accepted belief that there is a relationship between 
these two objects and furthermore, through Fermi surface reconstruction, stripe phases induce 
the formation of Fermi pockets (see reviews~\ci{Vojta,Vojta1} and references therein).

All the problems with understanding of experimental data result from the fact 
that the QPs, which determine the cuprate physics, have so far been considered as
fermions. A fermion statistics signature of these QPs has been detected
by observation of Fermi pockets, as constituents of a Fermi surface. However, the manifestation 
of fermions in a parent compound's Brillouin zone starts at low dopings in the form
of Fermi arcs. Then, upon evolution with doping, it acquires Fermi pockets, and, finally, 
after optimal doping, it becomes circle like, as for a homogenous Fermi gas~\ci{Fujita}.

However, the PG and thus the HTS cuprate physics is closely related with the MIC
mechanism discussed in the present paper. As seen from the list of the Coulomb two 
liquid model positions, the electric current properties, between the first and second 
critical dopings, are determined by two percolation channels. The first one is conducting
channel with percolation of NRs and thus particles inside of each NR; to this channel
belongs the bosonic insulator-bosonic metal-fermionic metal transitions. 

The second one belongs to the non-percolative, insulating channel. All fermions for 
dopings between the first and second critical concentrations are from this channel.
These charged fermions do not contribute in the PG current properties and physics,
therefore, we call them insulating fermions. The experimental information on these
fermions cannot be used for clarification of anything in the PG phase physics. 
Only from the second critical doping, when the percolation 
of NRs of these fermions together with fermions obtained in the percolative channel 
starts on, the charged fermions begin to play a role. Everything stated here for 
insulating fermions is actual for temperatures below the PG boundary $T^*$. Percolation
starts here from $T^*$.  

The information about the evolution of Fermi pockets with doping also finds understanding 
within the insulating fermions. Two critical doping 
values in the phase diagram temperature-doping of cuprates exactly 
relate with two maximal and minimal sizes of NRs, which means that distances between particles-fermions 
varies between these two NR sizes. If we denote the NR size at a given doping 
$x$ as $\xi_{coh}(x)$, then two limiting values for the Fermi pocket wave
vector are approximately determined as $k_{F1}\approx 2\pi/\xi_{coh}(x_{c2})$ and 
$k_{F2}\approx 2\pi/\xi_{coh}(x_{c1})$. Variation of these two values for the
wave vector with doping $x$ is given by formulas $k_{F1}\approx 2\pi/\xi_{coh}(x_{c2})$ 
and  $k_{F2}\approx 2\pi/\xi_{coh}(x)$.

Finally, we note that a Fermi surface in the form of disconnected
arcs or pockets is a signature of an insulator rather than a conductor. Because, for the
conductor, the wave vector of a charge carrier can be arbitrary inside of a Fermi 
surface: from zero value up to one of the finite Fermi wave vector~\ci{abrikosov}.

\section{Resistivity temperature upturn}
\label{sec3}  

The validity of the low-temperature (low-$T$) heat transport Wiedemann -- Franz law
(WFL): \be \kappa/(\sigma T)=L_0 \ , \lab{wf1} \ee where $L_0$ is the Lorentz ratio with
Sommerfeld's value $L_0=(\pi^2/3) (k_B/e)^2$ and $\kappa$ and $\sigma$ are the heat and electric
charge conductivities, respectively, has been perfectly proven for many
conventional metals~\ci{ashcroft}. We note that the $L_0$ value corresponds
to the three dimensional ($3D$) case. Simple calculation shows
that it can be also applied for the $2D$ case. An explanation of
the WFL for metals was based on the exploiting of the simple Drude
model for $\kappa$ and $\sigma$ involving Fermi liquid QPs~\ci{ashcroft}.
This model suggests the relation: \be \kappa =(1/2) c v l \lab{wf2} \ee
for $2D$ heat conductivity $\kappa$ (for the $3D$ case the factor
should be $1/3$) and \be \sigma=e^2 n \tau/m \lab{wf3} \ee for the
charge conductivity $\sigma$. Here $c$, $v$ and $l$ in Eq.~\re{wf2} are
the specific heat, mean velocity and mean free penetration length,
respectively, and $e$, $n$ and $\tau$ in Eq.~\re{wf3}  are the charge,
$2D$ density and mean lifetime, respectively, of Fermi liquid QPs
with mass $m$. For the low-$T$ limit the $T$ dependence of
$\kappa$ is entirely determined by the specific heat $c$. Since
the velocity $v$ (as a velocity on the Fermi surface) is a function of 
$n$ and the length $l\approx v\tau$ is also $T$ independent due to $\tau$ being
characterized by scattering of QPs with impurities~\ci{abrikosov} for 
these temperatures. For Fermi liquid QPs $c\sim T$ (see Ref.~\ci{abrikosov}), 
therefore, $\kappa\sim T$ and this leads to Eq.~\re{wf1} for WFL of that 
gas of QPs ($T$ independence of $\sigma$ is obvious from Eq.~\re{wf3}).

Hill et al. have reported on the first WFL measurement of electron-doped 
copper-oxide $Pr_{2-x}Ce_x CuO_4$ in the pioneer paper~\ci{hill}. 
They suppressed the HTS by the strong magnetic field and measured the low-$T$ 
dependence of the heat conductivity $\kappa$ for the PG QPs 
of cuprate. The observed $\kappa$ deviated from the normal linear 
$\kappa \sim T$ behavior into an anomalous $T^{3.6}$ with the decrease of
$T$. This behavior, called in the literature as downturn, has also been observed 
in the other hole-doped copper-oxides (see for references Refs.~\ci{Abdullaev0,Abdullaev01}).
The measurement of $\kappa$ was performed in these hole-doped copper-oxides likewise 
after suppression of the HTS by a strong magnetic field.

Smith et al.~\ci{Smith} have suggested by the phenomenological treatment that the 
downturn of $\kappa$ is a result of the decoupling of the heat carrying phonons 
and electrical charge carriers at low-$T$ ,while Hill et al.~\ci{hill} have noted 
that the downturn is a fundamental intrinsic property of copper-oxides.
The downturn has been widely accepted as a characteristic of non-Landau Fermi
liquid behavior. As the next evidence for the non-Fermi liquid QPs property
was recognized the fact that the cuprate Lorentz ratio was
significantly larger than Sommerfeld's value~\ci{proust}.

In our preceding paper (Ref.~\ci{Abdullaev0}) we have shown that these non-Landau Fermi
liquid behaviors of the low-$T$ heat conductivity and Lorentz ratio 
are results of the MIC property of the cuprate PG regions, in which the IGS
originates from a plasmon gas of 2D charged single bosons, pinned by a ferroelectric
field of parent compound atoms (see also previous section). In this section, 
we show that the downturn $T$ dependence carries the important information on 
the low-$T$ scale of the MIC charge conductivity or resistivity. It will be 
demonstrated that this scale of the charge conductivity is driven by that
of the specific heat. One should note that the physics of the MIC low-$T$ 
scale of copper-oxides was far from the clear understanding. 

As we argued in Ref.~\ci{Abdullaev0}, the magnetic field does not affect
on the energy of elementary excitations, i.e., plasmons, of gas of charged 
single bosons. The weak ferroelectric field effect of the cuprate atoms 
is also negligible for this energy. Therefore, we can use here results 
for the low-$T$ heat conductivity $\kappa$ and the specific heat $c$, 
obtained in ~\ci{Abdullaev0} for the above-mentioned gas of plasmons.
Namely, we assume that there exists a some downturn 
temperature $T_d^{\kappa}$ of $\kappa$, which separates the linear $\kappa\sim T$ 
dependence from the downturn $\kappa\sim T^4$ one, in the PG phase of
cuprate. Irrespective of the QP nature we consider, one may suppose the 
validity of WFL, Eq.~\re{wf1}, and Drude model, i.e., Eq.~\re{wf2} and 
Eq.~\re{wf3}, for these QPs. However, it is necessarily to replace in these 
formulas $L_0\rightarrow L$, since $L$ has the explicit doping dependence 
(see Refs.~\ci{Abdullaev0,proust}).   

On the other hand, the linear dependence $\kappa\sim
T$ for $T>T_d^{\kappa}$ resembles the case of conventional metals,
in which freely penetrating QPs are fermion-like and have $c\sim
T$ dependence for the specific heat. Additionally, for low-$T$ and
$T>T_d^{\kappa}$ we may again apply the above arguments for the $T$
independence of $v$ and $l$ for these QPs as well as conjecture
that the temperature dependent $\kappa$ arises from the $T$
dependence of the specific heat $c(T)$.  

We introduce the notations $c_1\sim T^4$ and $c_2\sim T$ for specific heats 
with $T^4$ and $T$ dependencies, respectively, and $\sigma_2$ for charge
conductivity, when $c_2\sim T$. Hence, the total specific heat $c$
transits from $c_1$ to $c_2$ as $T$ increases. The $c\sim T^4$
dependence is consistent with the function ${\cal S} \sim T^i$
with $i>1$ of an entropy (integral ${\cal S}=\int_{0}^{T} (c/T_1)
d \, T_1 \,$) observed in Ref.~\ci{loram1} for the insulating $Y Ba_2 Cu_3
O_{6+x}$ compound.

In the experimental papers on the WFL of cuprates, authors use the
residual value for the charge conductivity $\sigma_2$, which is the intercept
of $\sigma_2$, measured for $T\ge T_d^{\kappa}$ range, by extrapolation
to the $T\rightarrow 0$ limit. It was previously shown (see Refs.~\ci{abrikosov}) that the
typical low-$T$ $\sigma_2$ is nearly constant,
which is in accordance with the above assumption regarding the scattering
of freely penetrating QPs with impurities. Therefore, one assumes $\sigma=\sigma_2
\approx const$ in WFL, Eq.~\re{wf1}, in the experimental measurements in this
subject. We substitute the expressions for
$\kappa$, Eq.~\re{wf2}, and $\sigma_2$, Eq.~\re{wf3}, in the left hand side of
Eq.~\re{wf1}. Then recalling that $\kappa$ is expressed by the total
specific heat $c$, one obtains the formula \be \kappa/(\sigma_2
T)=cL/c_2 \  \lab{wf4} \ee for the measured WFL of cuprates, where
the constant $L$ is defined by the expression
$L=c_2mv^2/(2e^2nT)$. It is clear that $L$ is determined for the
$T\ge T_d^{\kappa}$ temperatures, for which the "metallic" QPs are
well defined. We cannot calculate the numerical value of $L$.
Since numerical data of empirical quantities, through which it is
expressed, are unknown. However, from the paper~\ci{proust} of Proust et
al. we know that $L$ is constant for fixed doping. Therefore,
Eq.~\re{wf4} demonstrates that for fixed doping, $T$ dependence of
WFL is provided by $c/c_2$ factor and vice versa, the observed
data on WFL determine the $T$ dependent $c/c_2$.

Let us suppose that the WFL for copper-oxides is temperature, $T$,
independent as for conventional metals and defined by the relation
$\kappa/(\sigma T)=L$, where $\sigma$ varies now in the whole
interval of $T$. Multiplying and dividing the left hand side of
this relation by $\sigma_2$ and using Eq.~\re{wf4} we obtain the
expression \be  \sigma=c \sigma_2/c_2 \ , \lab{wf5} \ee which is
the main result of the present section. Then, Eq.~\re{wf5} has two
asymptotic limits: $\sigma=\sigma_2$ for $T>T_d^{\kappa}$, when
$c=c_2$, and $\sigma\sim T^3\sigma_2$ for $T<T_d^{\kappa}$ and
$T\rightarrow 0$, when $c\sim T^4$. While for the first limit, the
charge conductivity $\sigma$ equals that of the conductor
$\sigma_2$ (since $\sigma_2=const\not=0$). For the second limit,
due to vanishing of $\sigma$ at $T\rightarrow 0$, we have the
conductor to insulator transition or the MIC. The important thing 
is the insulator charge conductivity has the $\sigma\sim T^3$ dependence.
This $T$ dependence for $\sigma$ may be used in the experiment for the
indication of the insulating state.

\begin{figure}
\begin{center}
\includegraphics[angle=0,width=8.5cm,scale=1.0]{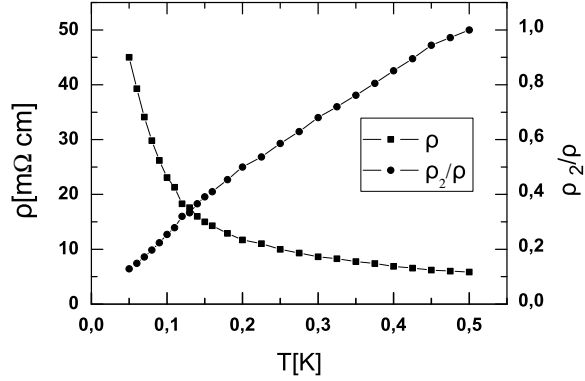}
\end{center}
\caption{The low-$T$ dependence of a resistivity $\rho$ for
$La_{2-x} Sr_x CuO_{4+\delta}$ copper-oxide with $x=0.06$ (from
Fig. 3 (a) of Ref.~\ci{hawthorn}) and constructed on the base of this $\rho$
our ratio $\rho_2/\rho$. It was assumed that metallic
$\rho_2=\rho(T=0.5K)$. } \lab{fig1}
\end{figure}

We may clarify the origin of the MIC crossover. For
this we extend the Drude formula, Eq.~\re{wf3}, for the charge
conductivity $\sigma$ to the all considered temperatures
(including the $T<T_d^{\kappa}$ interval) and define the
concentration of charge carriers $n$, which corresponds to the
$\sigma$. Let the "metallic" conductivity $\sigma_2$ be defined by
Eq.~\re{wf3} itself with the charge concentration $n_2$. Substituting
both Drude formulas for $\sigma$ and $\sigma_2$ in the Eq.~\re{wf5} 
we obtain $n=c n_2/c_2$. In the last equation, we have implied the
reasonable argument that other parameters of the gas of QPS
(except the concentration $n$) should not change in the
MIC. From this expression for $n$, we see
that $n=n_2$ for $T\ge T_d^{\kappa}$, when all charge carriers are
the freely penetrating QPs, and $n\sim T^3n_2\rightarrow 0$ for
$T<T_d^{\kappa}$ and $T\rightarrow 0$. The second limit of $n$
corresponds to vanishing of QPs. The decrease in the density of
the conducting carriers is a reason for the MIC of cuprates. It is 
also the origin of the decoupling of the heat-carrying thermal phonons
with these charge carriers~\ci{Smith}. 

The dependence for the concentration of carries, $n\sim T^3n_2$, as a function 
of temperature, $T$, and at fixed doping, $x$, can be experimentally measured 
for any copper-oxide by using the method, described, for example, in the 
Ref.~\ci{Laliberté}.

\begin{figure}
\begin{center}
\includegraphics[angle=0,width=8.5cm,scale=1.0]{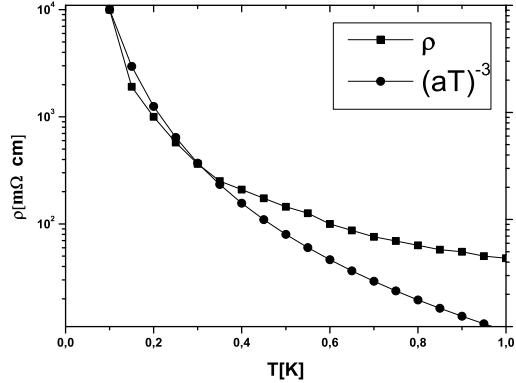}
\end{center}
\caption{The semi-log scale of the low-$T$ dependence of $\rho$
for $La_{2-x} Sr_x CuO_{4+\delta}$ copper-oxide with $x=0.05$
(measured without a magnetic field and below the first critical
doping) (the insert of Fig. 3 (a) of Ref.~\ci{hawthorn}) and our empirically
fitted function, $-3log(aT)$, with $a=0.464K^{-1}$. It is seen
that our function satisfactorily represents the non-linear part of
the observed $\rho$ at lowest $T$. } \lab{fig2}
\end{figure}

In Fig.1, we plot the low-$T$ dependence of a resistivity
($\rho=1/\sigma$) taken from Fig. 3 (a) of Ref.~\ci{hawthorn} for $La_{2-x}
Sr_x CuO_{4+\delta}$ copper-oxide with $x=0.06$. It is shown also
in this figure the ratio $\rho_2/\rho=\sigma/\sigma_2$, where we
assumed that $\rho_2=\rho(T=0.5K)$. The observed in~\ci{hawthorn} result for
$\rho$ corresponds to the insulating normal state of the compound
obtained by suppression of HTS by a
strong magnetic field. As seen from Fig.1, $\rho_2/\rho$ line has
the downturn behavior below $T=0.5K$ and an obvious non-linear
dependence for $T\le 0.15K$. This downturn relates to an
MIC. However, since the slope of the
depicted $\rho(T)$ is negative at $T=0.5K$, the $\rho_2/\rho=1$
assumption is approximate, and the copper-oxide is still in the
crossover state at this $T$. The downturn region of $\rho_2/\rho$
extends for higher temperatures.

In Fig.2, we compare in the semi-log scale the low-$T$ dependence
of $\rho(T)$ measured for purely insulating state $x=0.05$
(without magnetic field) of the same copper-oxide (the insert of
Fig. 3 (a) in Ref.~\ci{hawthorn}) with our empirically found $(aT)^{-3}$
function, where $a=0.464K^{-1}$. As is evident from this figure, our
function satisfactorily fits the non-linear part of the observed
$\rho(T)$ for the lowest temperatures. This non-linear $T$
dependence of $\rho(T)$ is also seen in Fig. 1 for lowest $T$. One
may assume both insulating states have the same origin and for
both, $T^{-3}$ dependence of $\rho(T)$ is valid.

At the end of this section, we note that the IGS of copper-oxides,
the origin of which is the plasmon gas of 2D charged single bosons, results
in the resistivity temperature upturn at the low-$T$ limit as at
strong magnetic field and $x>x_{c1}$, when the HTS is suppressed by 
this field, so and for purely insulating state with $x<x_{c1}$ and without field. 
This is in accordance with the point 10 of the Coulomb two liquid model 
positions and confirms the point 1, which means that, even though at 
$x<x_{c1}$ the electric charges and their NRs are not percolating, due to
plasmon origin of the IGS, the resistivity displays the upturn behaviour.
In principle, the last argument can also be applied for insulating fermions 
for dopings in the interval $x_{c1}\le x\le x_{c2}$. It seems that they
should give a similar upturn in the resistivity. However, the weight of their
concentration with respect to one of single bosons for the same dopings
is not high.  

\section{Nematicity and hidden magnetic order as intra PG pair charge and spin fluctuations}
\label{sec4}
First observed in $YBa_2Cu_3O_{6+x}$ (Y123) compound, using polarized elastic 
neutron diffraction~\ci{Fauque}, the  hidden magnetic order has been observed in 
three other copper oxide families: $HgBa_2CuO_{4+\delta}$ (Hg1201)~\ci{Li1,Li2}, 
$La_{2−x}Sr_xCuO_4$ (La214)~\ci{Baledent} and $Bi_2Sr_2CaCu_2O_{8+\delta}$ 
(Bi-2212)~\ci{Almeida} (see for references also~\ci{Mangin}). Since the size of 
detection was a few structural cells, it was called the intra-unit-cell (IUC) 
hidden magnetic order.  

These interesting experiments have revealed the existence of IUC objects in the PG phase 
with fluctuating spin components inside, which exactly cancel each other,
so that the total spin of the every object was zero. Despite the authors of Ref.~\ci{Fauque}
having interpreted the physics of both, total and intra, spins either by invoking of a pair of 
oppositely flowing intra structural cell charge loop-currents or of staggered spins in 
the same cell, the role of these objects in the physics of copper oxides was not 
understood~\ci{Norman}.

This role becomes unambiguously clear, if we connect these objects with the visualization 
of NRs which exhibit the energy gap in the STM experiment~\ci{Gomes}. Since, as NRs, they
exist in the PG region and disappear in the PG temperature boundary $T^*$~\ci{Fauque}. 
There is no doubt that their evolution with temperature will be the same as for NRs. 
While their evolution with doping (see Refs.~\ci{Fauque,Li1,Li2}) qualitatively 
coincides with that of the NRs, described in~\ci{Abdullaev1}. However, minimal size NRs 
are single bosons, therefore, the IUC hidden magnetic order objects are also 
single bosons and PG and HTS pairs. On the other hand, the spin fluctuation inside 
of the single boson was introduced in our original papers on the single boson
mechanism of HTS~\ci{Abdullaev3,Abdullaev2}. Furthermore, the increase of this
fluctuation with temperature below PG boundary $T^*$ and doping below the
second critical $x_{c2}$ and transformation of these single bosons into single fermions
in $T^*$ and $x_{c2}$ was physical origin of $T^*$ and $x_{c2}$.

The spatially intra rare charge density of each single boson 
allows one to understand the nature of the intra unit cell electronic nematicity, 
observed recently in the STM experiment~\ci{Lawler}, as the intra PG pair dynamic 
charge fluctuation. The strong ferrielectric crystal field of the parrent compound 
forms an atomic scale charge distribution within an individual NR (see for details
the Sec.~\ref{sec2}). 

The evolution of the dynamic charge fluctuation with doping resembles that of the 
IUC hidden magnetic order~\ci{Fujita} (Ref.~\ci{Fujita} contains a very thorough 
up-to-date list of references on the subject of IUC spin and charge fluctuations).

\section{Why ground state of $YBCO$ is Fermi liquid oscillating and of $Bi-2212$ is insulating?}
\label{sec5} 
The Fermi liquid like behavior of conducting QPs have been observed through the measurement of 
magnetoresistance oscillations at high magnetic fields in the underdome region of the 
temperature-doping phase diagram. The authors of these quantum oscillation experiments
have been succeeded in measuring a Fermi surface, which had a form of Fermi pockets. This effect 
was found for the YBCO family of cuprates for dopings up to optimal level (see the review~\ci{Vojta1} 
for references). 

However, the quantum oscillations  were not found in other copper-oxides, whose ground states were
insulating in the same magnetic fields (see Refs.~\ci{Ando,Fournier,Ono,Ando1,Boebinger}). There appears the question: Why the ground 
state of some copper-oxide compounds is Fermi liquid oscillating and of other cuprates is insulating
and can this discrepancy be understood within the framework of a single model?

If we look on the Table 1, we see that the critical doping for the $YBCO$ cuprate is $x_c=0.19$, while for
the $Bi-2212$ one is $x_c=0.28$. The lowest critical $x_c$ provides the percolation for the single fermion
NRs at the lower dopings, close to  $x_c$. Thus the $YBCO$ cuprate has the Fermi liquid oscillating 
ground state. While for $Bi-2212$, i.e., $Bi_2Sr_2CaCu_2O_{8+\delta}$ HTS compound, the single fermion 
NRs are not percolating for almost all $x$ between $x_{c1}$ and $x_{c2}$. Therefore, the ground state 
of $Bi-2212$ material in a strong magnetic field is insulating (see Ref.~\ci{Zavaritsky}). However, it 
is worthwhile to note here that the cuprate $LSCO$ with low critical doping $x_c=0.18$ 
(see Table 1) has, for unknown reasons to us, an insulating ground state 
(see Refs.~\ci{Ando,Fournier,Ono,Ando1,Boebinger}).    

\section{Interpretation of experimental MIC results 2016}
\label{sec6} 

The resistivity of cuprates at low temperatures and in high magnetic
fields revealed the appearance of an insulator-like upturn in its temperature
dependence~\ci{Ando,Fournier,Ono,Ando1,Boebinger}, which we already discussed in Sec.~\ref{sec3}. This upturn 
(MIC) behaviour is still considered puzzle in the literature.

Using high-field resistivity measurements on $La_{2-x}Sr_x CuO_4$, Laliberté et al.~\ci{Laliberté} 
have tried to experimentally show that this MIC upturn property is due to a drop in carrier 
density $n$ associated with the onset of the PG phase at a critical doping $x_c$. 
Namely, authors of Ref.~\ci{Laliberté} proposed that the upturns are quantitatively consistent 
with a drop from $n = 1 + x$ above $x_c$ to $n = x$ below $x_c$. Based on that idea, the explanations 
of the upturns for other copper-oxides are also given in~\ci{Laliberté}. Here, we note that 
this transition in carrier density $n$ dependence is the long term discussed in a literature issue 
(see~\ci{Laliberté} for references).  

First, we start with mentioning (see Sec.~\ref{sec2}) that all electric current properties
of cuprates should be considered withing the conducting percolation channel. Therefore, the 
resistivity temperature upturn is a result of this channel, more exactly, as we demonstrated 
in Sec.~\ref{sec3}, of the plasmon gas of 2D charged single bosons. No charged single fermions
below of critical doping $x_c$ play a role. Their NRs are not percolating, thus these fermions
are insulating. 

Secondly, one needs to make clear the definition of a carrier density. Because it does not simply
indicate the fraction of charge $x$ per $Cu$ atom. As we already discussed in previous sections, 
only the NR, i.e. the PG pair, carries the charge $n$ in the normal state. And for doping $x$ 
between critical $x_{c1}$ and  $x_{c2}$ we have $n=1+f(x)$, where the charge of boson is 1 and the
charge of fermion is $f(x)$. Here, the function $f(x)$ is phenomenological and taken from
the STM experimental data for NRs (see Table 1 for Bi-2212). Therefore, the fraction $f(x)$ 
indicates the charge of fermion inside of PG pair. At the approach of $x$ to $x_{c2}$, or 
for PG pair size approaching that of $2a$, the charge of fermion tends to 1 and thus each $Cu$ atom contains one 
doped fermion-hole. This is reason why at critical doping $x_c$ in Ref.~\ci{Laliberté} the charge
density adds 1 or becomes $n=1+x$. 

Laliberté et al. claim in the Ref.~\ci{Laliberté} that the transition $n = 1 + x$ into 
$n = x$ for the carrier density $n$ has a sharp, like in a second order phase transition,
form and it occurs exactly at the critical doping $x_c$. However, the original Refs.~\ci{Ando,Fournier,Ono,Ando1,Boebinger}
result on the upturns was MIC, which means a soft gradual transition, and it occured according
to~\ci{Ando,Fournier,Ono,Ando1,Boebinger} at "near optimal doping", i.e., below critical doping $x_c$. Furthermore,
Boebinger et al. demonstrated in the Ref.~\ci{Boebinger} that this MIC temperature-doping boundary
resembles the PG boundary $T^*$ but located below $T^*$. This MIC boundary 
approximately describes by Eq.~\re{wf7a} and corresponds to bosonic insulator-bosonic metal
transition boundary, as was discussed in our Ref.~\ci{Abdullaev0}, while the total PG phase MIC
corresponds to bosonic insulator-bosonic metal-fermionic metal transitions~\ci{Abdullaev01}. 
The last transition occurs at $T^*$.

One needs to note at the end of this section that the fraction of dopings corresponding to single 
bosons is hidden from the Luttinger sum rule investigations of copper-oxides. This rule 
investigates only fermions, however, there is a variety of publications, which supports or violates 
the Luttinger sum rule for these materials. Refs.~\ci{Mei,Yoshida}, for example, violates this rule.

\begin{table}[tb]
\begin {center}
\begin{tabular}{|c|c|c|c|c|c|} \hline
    $x/x_c$                                &  0.2    &  0.4   &  0.6   &  0.8   &  1.0     \\ \hline                                  
    $LSCO, x_c=0.18$                       &         &        &        &        &          \\
    $  n(x/x_c)$                           &   0.04  &  0.07  &  0.1   & 0.14   &  1.18    \\ \hline  
    $YBCO, x_c=0.19$                       &         &        &        &        &          \\
    $  n(x/x_c)$                           &   0.04  &  0.08  &  0.11  & 0.15   &  1.4$\pm$0.4   \\ \hline
    $Nd-LSCO, x_c=0.235$                   &         &        &        &        &          \\
    $  n(x/x_c)$                           &   0.05  &  0.09  &  0.14  & 0.188  &  1.3$\pm$0.4   \\ \hline
    $Bi-2212, x_c=0.28$                    & (0.06)  &(0.112) &(0.168) &(0.224) &  (1.3$\pm$0.4)    \\
    $ f(x)= n(x/x_c)-1$                    &  -0.103 & 0.170  &  0.341 & 0.524  &  0.939   \\ \hline
\end{tabular}
\end{center}
\vskip .5cm \caption{Carrier density as function of normalized 
to $x_c$ doping for $LSCO=La_{2-x}Sr_xCuO_4, YBCO=YBa_2Cu_3O_y$ and $Nd-LSCO=La_{1.6-x}Nd_{0.4}Sr_xCuO_4$, 
taken from Ref.~\protect\ci{Laliberté}, and for fermion carrier density 
for $Bi-2212=Bi_2Sr_2CaCu_2O_{8+\delta}$, taken from Ref.~\protect\ci{Abdullaev1}. 
For the comparison, in the brackets we showed numerical values for $Bi-2212$, if
the function was measured in the experiment. 
} \vskip -.5cm
\lab{tab-1}
\end{table}

\section{Conclusion}
\label{sec7}
In conclusion, we have found the understanding of the nematicity, 
IGS, nematicity to stripe phase transition, Fermi pockets evolution, and
resistivity temperature upturn, as to be MIC 
phenomena for the PG region of cuprates. While in our initial paper, we have found an understanding 
of the observed heat conductivity downturn, anomalous Lorentz ratio, insulator resistivity 
boundary, nonlinear entropy as manifestation of the same MIC. The observed nematicity and hidden 
magnetic order have been related to the PG pair intra charge and spin fluctuations. Probably, 
we could understand the answer on the question: Why ground state of $YBCO$ is Fermi liquid oscillating 
and of $Bi-2212$ is insulating? We have also analyzed the physics of the recently observed MIC result 
of Laliberté et al.~\ci{Laliberté} and explained the long term discussed in the literature 
transition of the electric charge density from doping to doping+1 dependence at the critical 
doping. We predict that at the upturns this density should have a temperature dependence of
$n\sim T^3n_2$ for $T\rightarrow 0$, where $n_2$ is density for dopings close to critical level. 
We understood that the upturns before and above first critical doping have the same nature.
We found understanding of all above mentioned phenomena within the PG pair physics.

\section*{Acknowledgements}

Authors B. Abdullaev, D.B. Abdullaev and M.M. Musakhanov acknowledge the support of the research 
by the Volkswagen Foundation of Germany and C. -H. Park by the National Research Foundation
of Korea (NRF) grant funded by the Korean Government (2015M3D1A1070639).

\end{document}